# Extending the dynamic strain sensing rang of ϕ−OTDR with frequency modulation pulse and frequency interrogation

JINGDONG ZHANG,[1,2] HAOTING WU,[1] JINGSHENG HUANG,[1] HUA ZHENG,[1] DANQI FENG,[1] GUOLU YIN,[1] TAO ZHU,[1,3]

[1]*Key Laboratory of Optoelectronic Technology & Systems (Ministry of Education), Chongqing University, Chongqing 400044, China*
[2]*Corresponding author: zjd@cqu.edu.cn*
[3]*Corresponding author: zhutao@cqu.edu.cn*

Nov. 19, 2019

**We propose and experimentally demonstrate a technique to extend the dynamic sensing range of phase−sensitive optical time−domain reflectometry (φ−OTDR) system based on the frequency interrogation. Benefitting from the range−Doppler coupling feature, the frequency modulation pulse is capable of measuring the frequency shift induced by the dynamic strain, thus the large dynamic strain can be recovered. The performance of the proposed method is experimentally evaluated by comparing it with phase unwrapping. The strain sensing rang can at least be increased by factor of hundreds, and fast dynamic strain with peak to peak 130 µε at vibration frequency of 20 kHz is measured.**

The phase−sensitive optical time−domain reflectometry (φ−OTDR) is one of the most effective way for distributed acoustic and vibration sensing (DAS/DVS), which is growing demanded in fields like geological survey and structural health monitoring [1-4]. In φ−OTDR, multiple vibration points along the long range sensing fiber can be sampled simultaneously through demodulating the phase changes of the backscattering light generated by the coherent light pulse. However, since the phase unwrapping process must be performed to obtain the continuous dynamic strain signal induced by the vibration[5], the absolute difference of adjacent phase changes of one vibration point needs to be less than to π rad. Thus, the amplitude of the vibration, especially for high frequency signal, is restricted, which limits the applications of φ−OTDR in case where the large dynamic strain measurement is required.

In this study, a new concept for the distributed sensing of large dynamic strain with φ−OTDR system is proposed and demonstrated. The measurement protocol relies on the frequency change interrogation ability that is provided by the frequency modulation pulse[6]. Apart from the phase of the backscattering light, the laser frequency change induced by the vibration is also demodulated by evaluating the temporal shift of the backscattering trace, thus, the rapid and large strain change can be measured. The quantitative relationship between the laser pulse modulation parameters and the strain change is derived in this paper. The sensitivity and the measurement accuracy of this proposed method are also analysed and tested experimentally, which reveals the dynamic strain sensing rang can be increased by factor of hundreds.

The dynamic stain sensing range in phase interrogation based φ−OTDR system is limited by phase unwrapping, and is affected by parameters like sampling frequency, phase demodulation noise, and signal frequency. Taking the sinusoidal signal as an example, the maximum detectable amplitude of the phase difference induced by dynamic strain in φ−OTDR system can be estimated as

$$\begin{cases} 2A\sin\left(\pi\dfrac{f}{f_s}\right) < \pi - 6\delta_p, & f < f_s \\ 2A < \pi - 6\delta_p, & f \geq f_s \end{cases}, \qquad (1)$$

where $A$ and $f$ are the amplitude and frequency of the sinusoidal phase signal, $f_s$ is the sampling rate or pulse repetition rate of the φ−OTDR system, and the $\delta_p$ is the standard deviation of the phase demodulation error. Function (1) is an empirical formula to estimate the detectable amplitude and verified by Monte Carlo simulation with the assumption that the phase demodulation error obeys a zero mean Gaussian distribution.

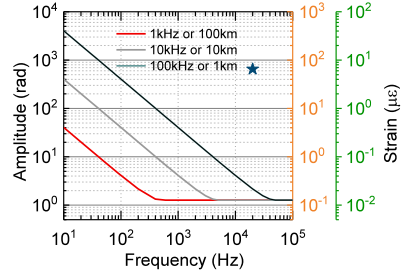

Fig. 1. The maximum detectable phase amplitude with vibration frequency in φ−OTDR system.

In Fig. 1, the detectable phase amplitude $A_{max}$ of the system is shown with $\delta_p$=0.1 rad, reporting the $A_{max}$ as a function of dynamic signal frequency $f$, as well as the sampling rate $f_s$ or sensing length $L$. The detectable amplitude $A_{max}$ is roughly in inverse proportion to the signal frequency $f$ and the sensing length $L$, and reduces to $(\pi-6\delta_p)/2$ when $f$ increases to $f_s$. We also calculated the corresponding detectable strain amplitudes for 1.1 m and 11.3 m phase gage lengths, which are labeled as yellow and greed axes on the right side, respectively. For a typical system with 10 km sensing length and 11.3 m gage length, the maximum detectable amplitude of the applied dynamic strain is 1με at 40.5 Hz and drops to 100 nε at 405.6 Hz The limitation in the dynamic strain sensing range hinders the application of φ−OTDR in field where strong dynamic strain exists.

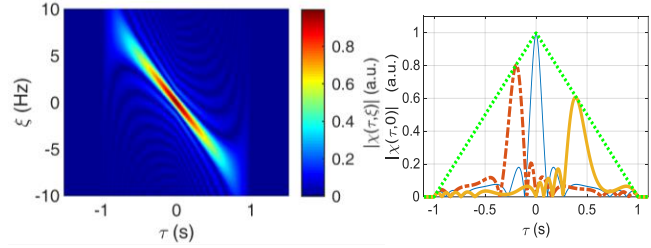

Fig. 2. (a) The ambiguity function of the LFM pulse with $B$=10 and $T$=1. (b) Zero Doppler cut and Doppler−shifted cut of the AF. Blue, orange, and yellow line are the zero Doppler cut, 2 Hz and −4 Hz Doppler−shifted cut of the AF. The green dot line is the zero Doppler cut of the AF of LFM pulse with $B$=0 and $T$=1.

With the assistance of matched filtering, the linear frequency modulation (LFM) pulse enables separate control of pulse energy (through pulse duration) and range resolution (through the pulse sweep bandwidth). Meanwhile, the LFM pulse provides the ability of the Doppler shift measurement, which can be used to evaluate the pulse center frequency shift induced by large dynamic strain in φ−OTDR. The ambiguity function (AF) is an analytical tool to examining resolution, side lobe, ambiguities in both range and Doppler, as well as range−Doppler coupling of a given waveform paired with its matched filter. The ambiguity function for a waveform with complex envelope of $x(t)$ is defined as

$$\left|\chi(\tau,\xi)\right|=\int_{-\infty}^{+\infty}x(t)x^{*}(t-\tau)e^{j2\pi\xi t}dt , \qquad (2)$$

where $\tau$ represents the ranging axis and $\xi$ stands for the Doppler or frequency shift axis. For an up−chirp LFM pulse with bandwidth $B$=10 Hz and pulse time−width $T$=1 s, the corresponding AF is shown in Fig. 2(a), which is shaped as a skewed triangular ridge and reveals the time−frequency coupling phenomenon of LFM pulse. The zero Doppler cut of the AF is just the matched filter output when there is no Doppler frequency mismatch, as shown as blue line in Fig. 2(b). When there is a Doppler mismatch, the peak of the matched filter output will be shifted proportionally to the frequency shift $\Delta\xi$ and occur at

$$\Delta\tau=-\frac{\Delta\xi T}{B} . \qquad (3)$$

Meanwhile, the amplitude of the peak will also be reduced, as illustrated in Fig. 2(b). According the time−frequency coupling phenomenon of LFM pulse, the large dynamic strain can be measured by demodulating the center frequency shift of the LFM pulse according the temporal shift of the backscattering trace.

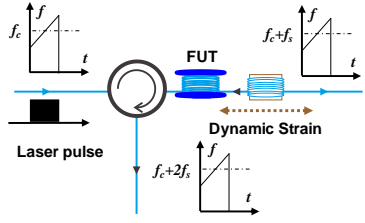

Fig. 3. The frequency modulation of the LFM pulse by dynamic strain in φ–OTDR system

Figure 3 schematically depicts the dynamic strain induced frequency modulation of the LFM pulse in φ–OTDR system. The incident LFM can be expressed as

$$x_l(t) = rect(\frac{t}{T})\exp\left(j2\pi f_c t + \frac{j\pi B}{T}t^2\right),\qquad(4)$$

where $f_c$ is the center frequency of the LFM pulse. When the pulse pass through the dynamic strain region, the dynamic strain induced phase modulation $\varphi(t)$ is

$$\varphi(t) = \frac{2\pi(n+C_\varepsilon)\varepsilon(t)L}{\lambda},\qquad(5)$$

in which $n$ is the effective refractive index of the sensing fiber, $C_\varepsilon$ is the strain parameter of refractive index, $L$ is the length of the dynamic strain region, $\lambda$ is the wavelength of the pulse and $\varepsilon(t)$ is the applied dynamic strain. Thus, the dynamic strain induced frequency modulation $f_s(t)$ is expressed as

$$f_s(t) = \frac{d(\varphi(t))}{2\pi dt} = \frac{L(n+C_\varepsilon)}{\lambda}\frac{d\varepsilon(t)}{dt}.\qquad(6)$$

As for the backscattering signal, the total dynamic strain induced frequency modulation $\Delta\xi(t)$ is given by the double of $f_s(t)$ for the backscattering pulse passing through the strain segment twice during its roundtrip, which is $\Delta\xi(t) = 2f_s(t)$. As a result, the matched filter output of the backscattering trace experiences a distance shift $\Delta d$ in the rear of the strain segment and

$$\Delta d(t) = \frac{c\Delta\tau(t)}{2n} = -\frac{cTL(n+C_\varepsilon)}{\lambda Bn}\frac{d\varepsilon(t)}{dt}.\qquad(7)$$

Equation (7) shows that the distance shift in the rear of strain section $\Delta d$ is proportional to the differential of applied dynamic strain. To characterize $\Delta d(t)$, on another front, the time–delay estimation algorithms like cross correlation can be applied. The achievable sensitivity and sensing range of this method are mainly determined by the $T$ and $B$ of the LFM pulse, the sampling rate, and the length of segment to evaluate the distance shift in time–delay estimation. Note that the frequency interrogation of the backscattering signal is based on the amplitude of the matched filtering traces, nevertheless, the phase traces may still be used to demodulation the small dynamic strain.

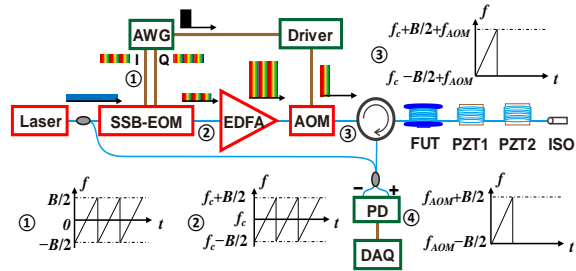

Fig. 4. Experimental setup of the proposed large dynamic strain distributed sensing system. AWG, arbitrary waveform generator; SSB–EOM, single sideband electro–optic modulator; EDFA, erbium–doped fiber amplifier; AOM, acousto–optic modulator; FUT, fiber under–test; PZT, piezoelectric transducer; ISO, isolator; PD, photo detector; DAQ, data acquisition card.

Figure 1 shows the configuration of the proposed dynamic strain sensing system. A 1550 nm narrow line–width laser (NKT, E15) is used as the light source and split into two branches by a fiber coupler. The upper branch is modulated by linear frequency sweeping signals of an AWG via a SSB–EOM. The AWG working at repetition mode and the RF output is series of orthogonal LFM wave with frequency range of –$B/2$ to $B/2$, as show in the inset, and the single frequency laser center at $f_c$ is modulated to a continuous LFM laser at range of $f_c+B/2$ to $f_c+B/2$. An EDFA is set at the front of AOM (G&H, Fiber−Q), which is used to compensate the insert loss of SSB−EOM and AOM, and boost the optical power before the light is launched to the FUT. The AOM is synchronous controlled by the AWG and chops the continuous LFM laser to LFM laser pulse with length of $T$, meanwhile, the pulse center frequency undergoes frequency up−shift by $f_{AOM}$, in which $f_{AOM}$ is the driving frequency of AOM and $f_{AOM}$=300 MHz in our system. At the real end of the FUT, the reflection light from the fiber end is blocked by an ISO. The backscattering light from the FUT interferes with the lower branch reference light, and the beat signal is situated in a frequency band between $f_{AOM}-B/2$ to $f_{AOM}+B/2$, if the dynamic stain induced frequency shift is neglected. The beat signal is detected by a balanced PD (Thorlabs, PDB480C) and digitized by the DAQ which is synchronized and triggered by AWG. We set the EDFA in the front of the AOM because the LFM lowers the power requirement of the pulse. Significantly, the pulse distortion and ASE (amplified spontaneous emission) noise induced by the transient effect of EDFA are also avoided, comparing with chopping the light before amplification.

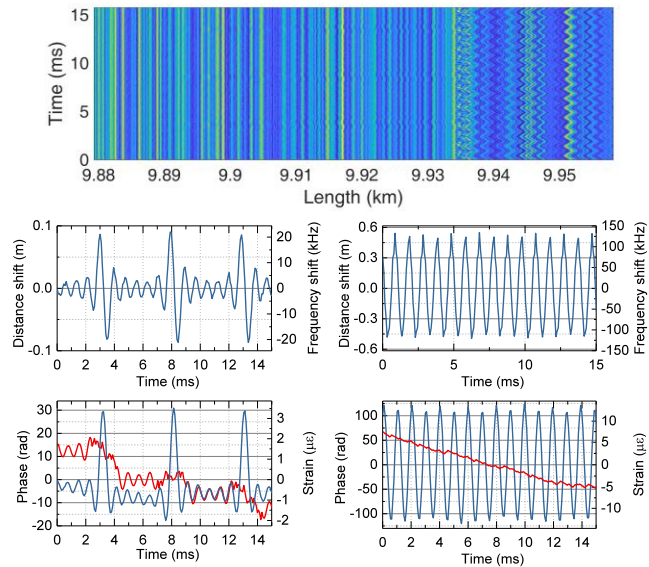

Fig. 5. (a) Amplitude trace map demodulated with matched filter when signals are applied on the PZTs. (b-c) The distance shifts of the backscattering trace at PZT1 (b) and PZT2 (c). The right axes are the demodulated frequency changes. (d-e) The blue lines are the phase signals demodulated by the frequency changes, and the red lines are demodulated by the phase trace map. The right axes are the calculated strain.

In the experiment, two PZT tubes are placed at the far end of the 10 km FUT with 30 m spacing, and two sections of fiber with length of 1 m are wrapped around and glued on the PZT tubes. Repeated Sinc function signal and sinusoidal signal are applied to the PZTs after amplified by the high voltage amplifier (TEGAM, 2350), in which the main lobe width of the Sinc function and the frequency of the sinusoidal signal are 769 μs and 1 kHz, respectively. We first experimented using a LFM laser pulsed with $B$=550 MHz and $T$=20 μs. Figure 5 shows the demodulated amplitude trace map at the end section of FUT when signals are applied on the PZTs. Clearly, the amplitude traces at the near region of PZT1 keep steady, however the traces at the far region of both PZT1 and PZT2 present distance shift for the light frequency is modulated by the dynamic strain. The Fig. 5 (b) and (c) show the distance shifts of the amplified traces at PZT1 and PZT2. The frequency changes of light induced by the dynamic strain at PZT1 and PZT2 are calculated using the range−Doppler coupling relation of the LFM pulse according Eq. (3), and labeled on the right axes in Fig. 5 (b) and (c). Then, the phase changes at PZT1 and PZT2 can be calculated by integrating the frequency changes, and are provided as blue lines in Fig. 5 (d) and (e), respectively. As expected, the repeated Sinc function signal and sinusoidal signal are well demodulated, with the calculated dynamic strain labeled on the right side. For comparison, the phase wrapping signal are also shown in Fig. 5 (d) and (e) as marked as red lines. It can be seen that, in Fig. 5(d), the fiber undergo large dynamic strain at the periods of main lobes of Sinc function. Thus, the induced phase changes between adjacent sampling points are larger than π rad, and the phase change can not be demodulated correctly with the phase unwrapping algorithm. Nevertheless, the phase changes at the periods of side lobes of Sinc function are within π rad and can be well unwrapped. It also demonstrated that the phase demodulated by both the phase unwrapping and the frequency interrogation methods are in good agreement with each other at the side lobes periods, and the main lobes are well demodulated by frequency interrogation. Figure 5(e) shows the phase results of PZT2 with 1 KHz sinusoidal large dynamic strain. The phase amplitude is over 100 rad, thus can only be demodulated correctly by the frequency interrogation method. The calculated strain are labeled on the right axes of Fig. 5(d) and (e). The experiment results in Fig. 5 indicates the frequency interrogation method is capable for large dynamic strain demodulation.

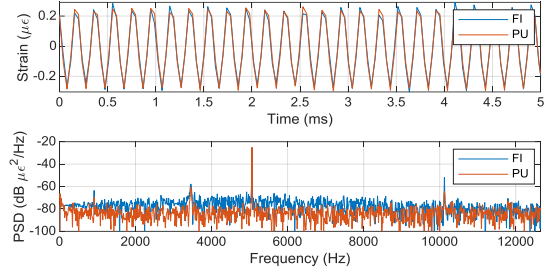

Fig. 6. Comparison of demodulation results by frequency interrogation and phase unwrapping. (a) Demodulated dynamic strain signals. (b) The power spectrum density (PSD) of the demodulated signals. FI, frequency interrogation method. PU, phase unwrapping method.

Next, the demodulation noises of the two methods are compared by using 1 km FUT. The repetition rate of the LFM pulse is increased to 25.3 kHz with the $B$ and $T$ of the pulse keeps unchanged. A 5 kHz sinusoidal signal is applied to the PZT2 and the demodulated dynamic signals, as well as their power spectrum density (PSD), are plotted in Fig. 6. It is obvious that the dynamic strain are well demodulated by both phase unwrapping and frequency interrogation methods. The noise floor of the phase unwrapping result, roughly situated at -80 dB µε$^2$/Hz, is 10 dB lower the noise floor of frequency interrogation result.

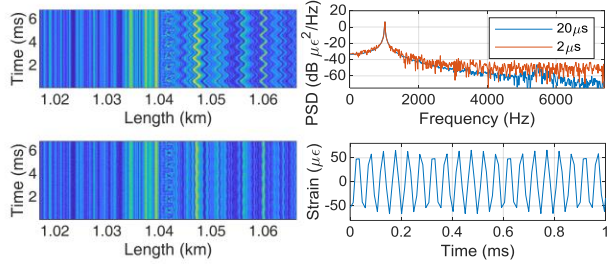

Fig. 7. The demodulation trace with 20 µs (a) and 2 µs (b) LFM pulse. The PSDs of the demodulated dynamic strain signals. (d)

Since the distance shift $\Delta d$ is proportional to the time–width $T$ of the LFM pulse according Eq. (7), the sensitivity of the frequency interrogation method is increase with $T$. We use a LFM pulse pair with $T$=20 µs and $T$=2 µs to analyze the influence of $T$ on sensitivity and demodulation noise. A 1 kHz sinusoidal signal is applied to the PZT2, and the repetition rate of the pulse pair is 14.8 kHz. The demodulated amplitude traces shown in Fig. 7(a) and (b) reveal that the $\Delta d$ is increased when longer LFM pulse is adopted, thus, the demodulation noise is lower for the 20 µs pulse, as shown in Fig. 7 (c). However, the detectable dynamic sensing range of the shorter pulse is larger for the dynamic strain induced frequency change is stable when the pulse path through the strain range. We drive the PZT2 at its resonant frequency with 20 kHz sinusoidal signal, therefore a high frequency dynamic strain is applied to the fiber. The repetition rate of the 2µs LFM pulse is set to 83.3 kHz, and the demodulation strain is shown in Fig. 8(d). The peak–to–peak value of the strain is over 130µε, as marked as a blue star in Fig. (1), which is over 100 times beyond the detectable range of phase unwrapping method. It also worth noting that the detectable range of the frequency interrogation method is irrelevant to the pulse repeating rate of the system.

In conclusion, we have proposed and experimentally demonstrated a new method for large dynamic strain sensing based on strain induced frequency shift interrogation. The key of the approach is that the range and Doppler coupling property of the LFM pulse. The proposed method is validated by comparing its demodulation result with phase unwrapping, and the detectable range can go beyond 130 µε at 20 kHz. The frequency interrogation method tremendously extends the dynamic strain sensing range of φ–OTDR, which makes φ–OTDR system suitable for various applications in real-world environments.